\documentclass[3p,times]{elsarticle}
\usepackage{ecrc}
\usepackage{epstopdf}
\volume{00}
\firstpage{1}
\journalname{Nuclear Physics A}
\runauth{Alberica Toia}
\jid{nupha}
\jnltitlelogo{Nuclear Physics A}
\usepackage{graphicx}
\usepackage{amsmath,amssymb}

\newcommand{\pt}           {\ensuremath{p_{\rm T}}}
\newcommand{\pp}           {pp}
\newcommand{\PbPb}         {\mbox{Pb--Pb}}
\newcommand{\pPb}          {\mbox{p--Pb}}

\newcommand{\pA}           {\mbox{p--A}}
\newcommand{\Npart}        {\ensuremath{N_\mathrm{part}}}
\newcommand{\Nparttar}     {\ensuremath{N_\mathrm{part}^\mathrm{target}}}

\newcommand{\Ncoll}        {\ensuremath{N_\mathrm{coll}}}

\newcommand{\qpa}          {\ensuremath{Q_{\rm pPb}}}
\newcommand{\rpa}          {\ensuremath{R_{\rm pPb}}}
\newcommand{\Nch}          {\ensuremath{N_\mathrm{ch}}}
\newcommand{\dNdeta}       {\ensuremath{\mathrm{d}N_\mathrm{ch}/\mathrm{d}\eta}}

\begin{document}

\begin{frontmatter}

\title{Centrality Dependence of Particle Production in p--A collisions measured by ALICE}

\author{Alberica Toia (for the ALICE\fnref{col1} Collaboration)}
\fntext[col1] {A list of members of the ALICE Collaboration and acknowledgements can be found at the end of this issue.}
\address{Goethe University Frankfurt, GSI Darmstadt}

\begin{abstract}
We present the centrality dependence of particle production in \pA\
collisions at $\sqrt{s_{NN}}$ = 5.02 TeV measured by the ALICE experiment,
including the pseudo-rapidity and transverse momentum spectra, with a
special emphasis on the event classification in centrality classes and
its implications for the interpretation of the nuclear effects.
\end{abstract}


\end{frontmatter}

%%\linenumbers

\section{Introduction}
Studies of nuclear effects in minimum bias (MB)
\pPb\ collisions for charged particles~\cite{alice_RpA_new}, heavy
flavor and jets show that the observed strong suppression in
\PbPb\ collisions is due to final state effects. Centrality dependent
measurements of the nuclear modification factor require the
determination of the average number of binary collisions, \Ncoll, for
each centrality class. Moreover, it has been recognized that the study
of \pPb\ collisions is interesting on its own, with several
measurements~\cite{Abelev:2012ola,Abelev:2013bla,Abelev:2013haa,ABELEV:2013wsa}
of particle production in the low and intermediate \pt\ region that
can not be explained by an incoherent superposition of
\pp\ collisions, but rather call for coherent and collective effects.

To determine centrality in ALICE we use as many detectors as
possible~\cite{AlicePerf}, in various rapidity
regions~\cite{Alice:Centrality}. Particle production measured by
detectors at mid-rapidity can be modeled with a negative binomial
distribution (NBD), while the zero-degree energy measures the slow
nucleons emitted in the nucleus fragmentation process, which we model
with a Slow Nucleon Model (SNM) \cite{AToia:2014}. These models (NBD
and SNM) are coupled to a \pPb\ Glauber MC and \Ncoll\ are obtained
for each centrality class determined by slicing the experimental
distribution in percentiles of the hadronic cross-section. The
\Ncoll\ values are similar for different estimators within the
systematic error from the Glauber parameters and a MC closure test
with HIJING.

However, in order to use these \Ncoll\ values in a $\rpa (\pt , {\rm
  cent}) = \frac{{\rm d}N^{\rm pPb}_{\rm cent}/{\rm d}\pt} { \langle
  N_{\rm coll}^{\rm cent} \rangle {\rm d}N^{\rm pp}/ {\rm d}\pt}$
calculation, one needs to take into account the bias arising when
sampling the \pA\ events in centrality classes.  In \pPb\ collisions,
the range of multiplicities used to select a centrality class is of
similar magnitude as the fluctuations, with the consequence that a
centrality selection based on multiplicity may select a biased sample
of nucleon-nucleon collisions. In essence, by selecting high (low)
multiplicity one chooses not only large (small) average \Npart\, but
also positive (negative) multiplicity fluctuations. These fluctuations
are partly related to qualitatively different types of collisions,
described in all recent Monte Carlo generators by fluctuations of the
number of particle sources via multi-parton interaction.  Concerning
the nuclear modification factor other types of bias have been
discussed: the jet-veto effect, due to the trivial correlation between
the centrality estimator and the presence of a high-\pt\ particles in
the event; the geometric bias, resulting from the mean impact
parameter between nucleons rising for most peripheral events.  Studies
of particle production and centrality determination have already been
presented by ALICE \cite{AToia:2014}; here we focus on the results
obtained with a new approach, described in the following sections.

\section{The hybrid method}\label{sec:hybrid}  
The hybrid method aims at providing a data driven and
unbiased centrality determination.  We give priority to a centrality
selection with minimal bias and, therefore, use the signal in the Zero
Degree Calorimeter (ZNA). In this case we cannot establish a direct
connection to the collision geometry but we can study the correlation
of two or more observables that are causally disconnected after the
collision, e.g because they are well separated in rapidity.

\begin{figure}
\begin{center}
\includegraphics*[width=0.495\textwidth]{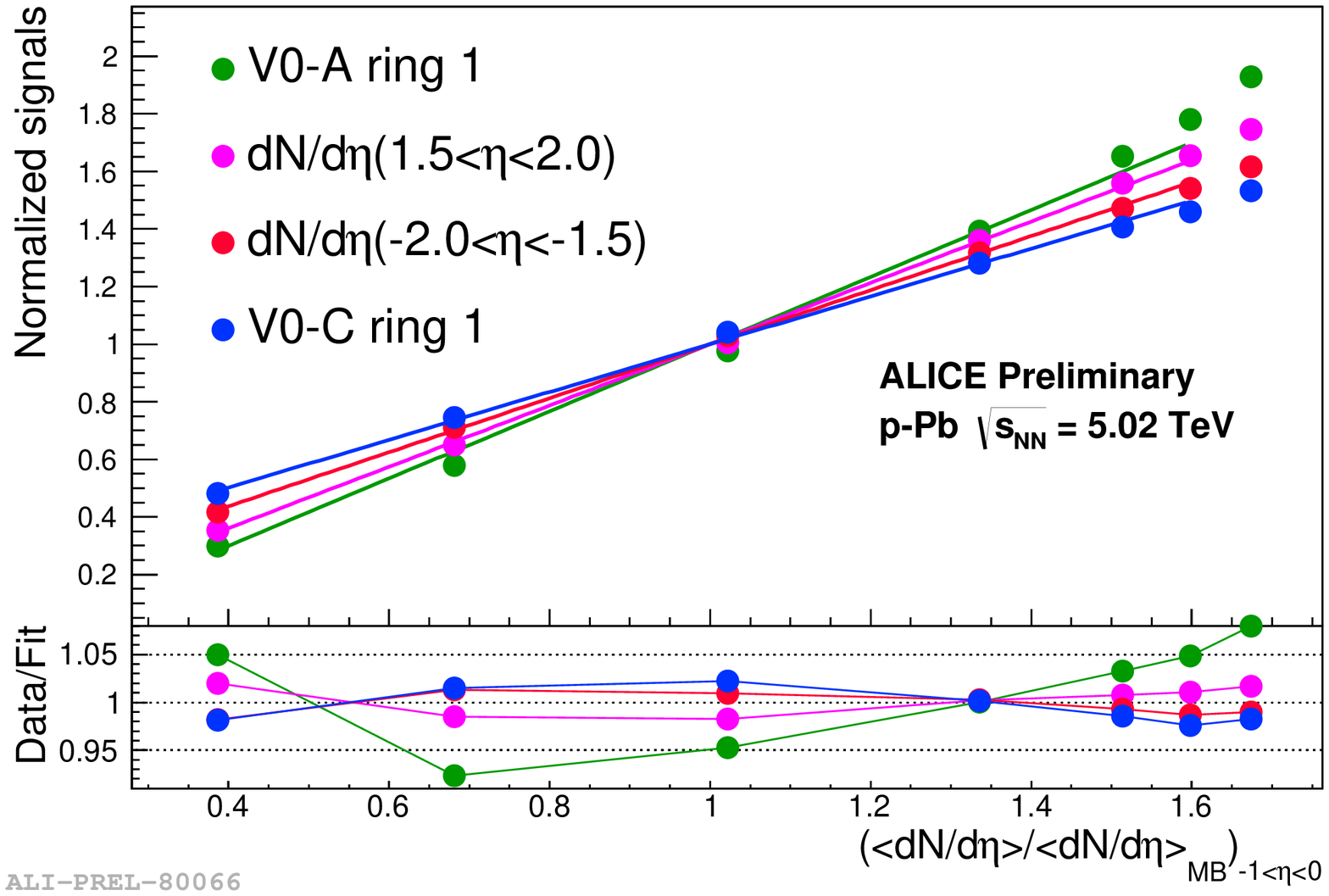}
\includegraphics*[width=0.35\textwidth]{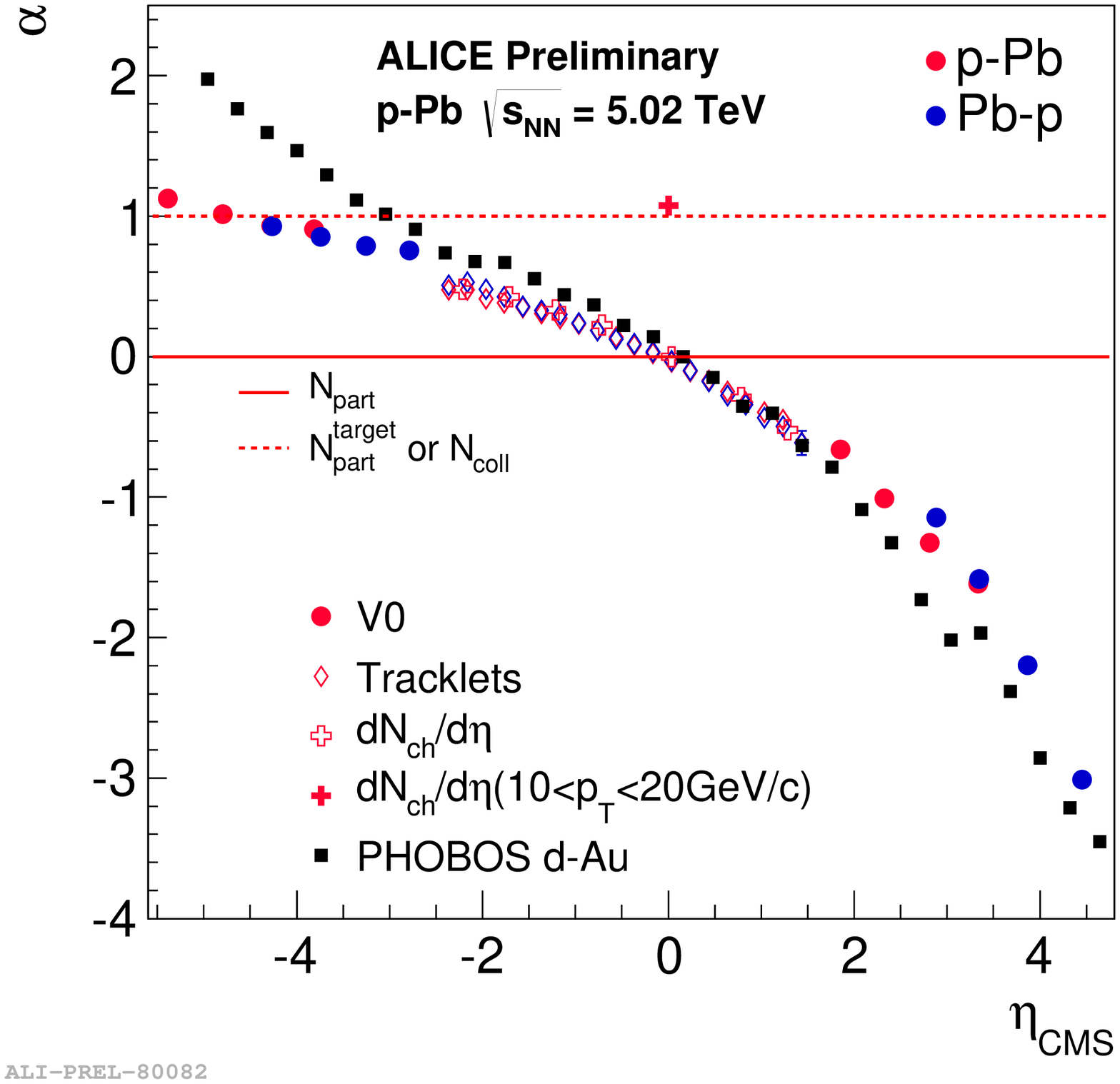}
\caption{Left: Normalized signal from various observables versus the
  normalized charged-particle density, fit with a lixsnear function of
  \Npart. Right: Results from the fits as a function of the
  pseudorapidity covered by the various observables. The red
  horizontal lines indicate the ideal \Npart\ and \Ncoll\ geometrical
  scalings. The most central point is excluded from the fit, to avoid
  pile-up effects.}
\label{fig:hybrid}
\end{center}
\end{figure}

Charged particle multiplicity is dominated by soft particles while
hard processes are expected to scale with \Ncoll. In centrality
classes selected by ZNA, we study the dependence of various
observables in different $\eta$ and \pt\ regions on the charged
particle density at mid-rapidity.  In order to compare different
observables on the same scale and also to neglect efficiency and
acceptance, we normalize the values in different classes by the
corresponding MB value.  The correlation of the signals to the
mid-rapidity particle density (Fig.\ref{fig:hybrid} left) exhibits a
clear dependence on the rapidity. The slope of the signals with
\dNdeta\ decreases towards the proton direction (C-side).  In the
Wounded Nucleon Model, \Npart\ is expressed in terms of target and
projectile participants. The particle density at mid-rapidity is
proportional to \Npart, whereas at higher rapidities the model
predicts a dependence on a linear combination of the number of target
and projectile participants with coefficients which depend on the
rapidity. Close to Pb-rapidity a linear wounded target nucleon scaling
(\Nparttar\ = \Npart\ - 1) is expected.

In order to further quantify the trends of the observables and to
relate them with geometrical quantities, such as \Npart, one can adopt
the WNM model and make the assumption that the charged particles
density at mid-rapidity is proportional to \Npart\ and relate the
other observables to \Npart\ assuming linear dependence, parameterized
with \Npart\ - $\alpha$. The results for $\alpha$
(Fig.\ref{fig:hybrid} right) indicate a monotonic change of the
scaling with rapidity. The red horizontal lines show the ideal
geometrical scalings. In Pb-going direction (negative $\eta_{\rm CMS}$
in the figure) the values of $\alpha$ reach the ones obtained for
charged-particle production at high-\pt. In contrast, in the
proton-going direction, $\alpha$ is much lower, indicating strong
suppression of the charged-particle production with centrality with
respect to \Npart-scaling. The correlation between the ZDC energy and
any variable in the central part shows unambiguously the connection of
these observables to geometry. Our data are overlaid with the
corresponding fit parameters derived from PHOBOS in d--Au collisions
at $\sqrt{s_{NN}}$ = 200 GeV. The comparison shows a good agreement
over a wide $\eta$ range, with some deviations at large negative
pseudorapidity. In particular, the $\eta$ region covered by the
innermost ring of the VZERO-A detector corresponds to the target
fragmentation region where extended longitudinal scaling was observed
at RHIC~\cite{Back:2004mr}.

Exploiting the findings from the correlation analysis described, we
make use of observables that are expected to scale as a linear
function of \Ncoll\ or \Npart, to calculate \Ncoll:
\begin{itemize}
\item $\Ncoll^{\rm mult}$: the multiplicity at mid-rapidity
proportional to the \Npart;
\item $\Ncoll^{\rm Pb-side}$: the target-going multiplicity
proportional to \Nparttar;
\item $\Ncoll^{\rm high-\pt}$: the yield of high-\pt particles at
mid-rapidity is proportional to \Ncoll.  
\end{itemize}
These scalings can be used as an ansatz to calculate \Ncoll, rescaling
the MB value $\Ncoll^{\rm MB}$ by the ratio of the normalized signals
to the MB one.  We therefore obtain 3 sets of values of \Ncoll, whose relative
difference does not exceed 10\%. This confirms the consistency of the
used assumptions, although it does not constitute a proof that any or
all of the assumptions are valid.

\section{Physics Results}
\subsection{Nuclear Modification Factors}
\begin{figure}[t!f]
 \centering
\includegraphics[width=0.4\textwidth]{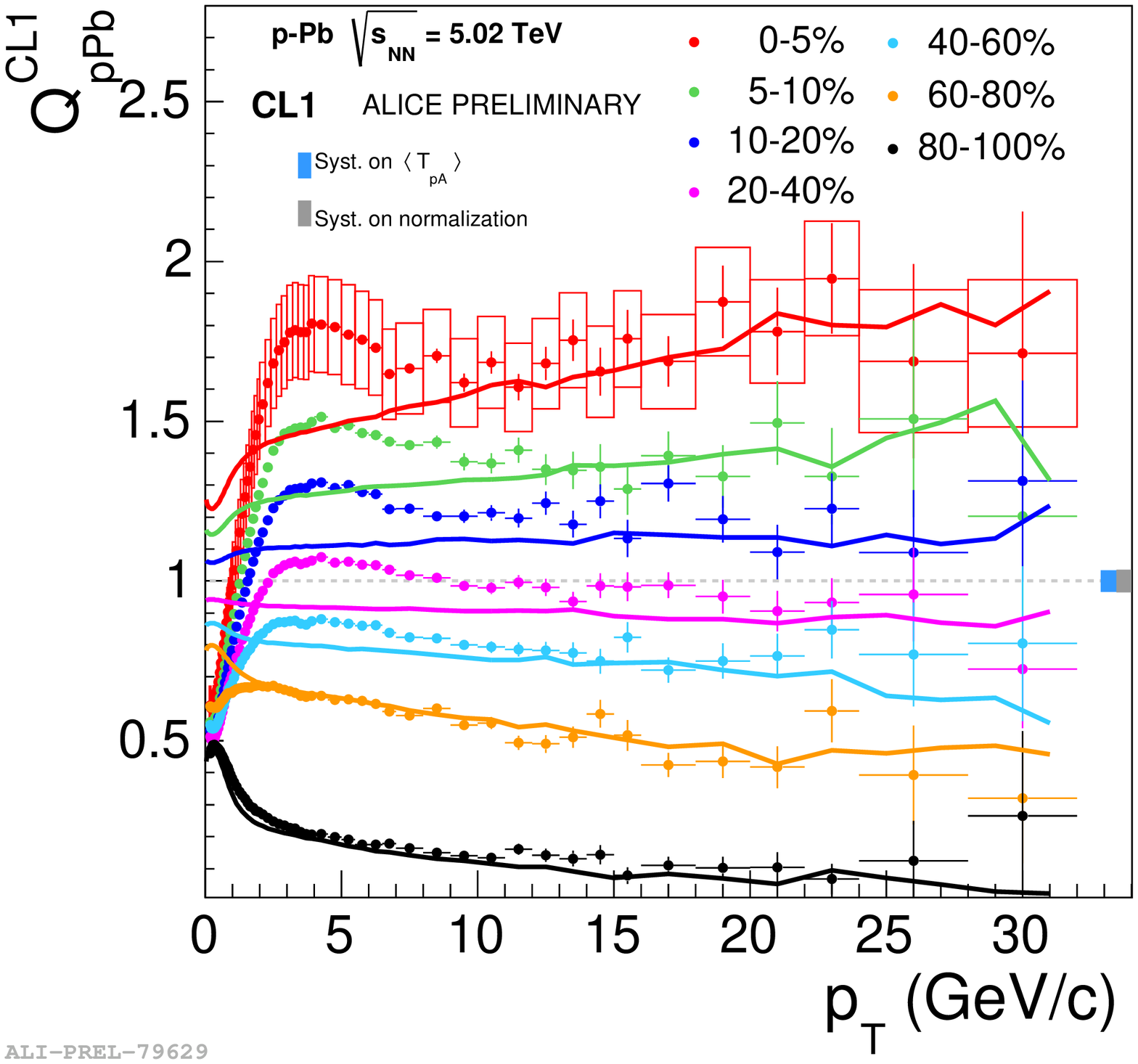}
\includegraphics[width=0.4\textwidth]{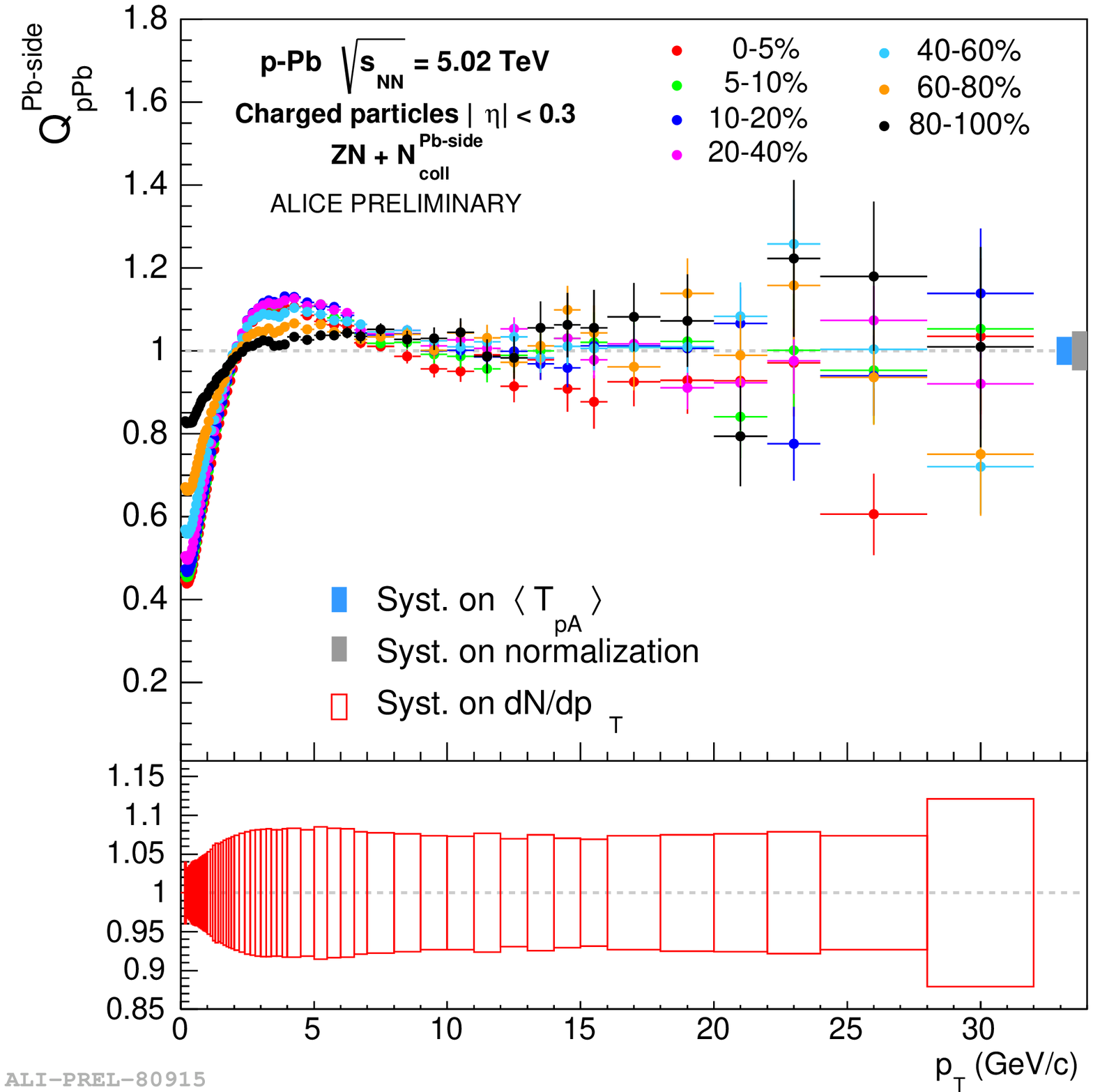}
 \caption{\qpa\ calculated with CL1 estimator (left), the lines are
   the G-PYTHIA calculations; with the hybrid method (right), spectra
   are measured in ZNA-classes and \Ncoll\ are obtained with the
   assumption that forward multiplicity is proportional to \Nparttar.
  \label{fig:QpAhybrid}}
\end{figure}

As already discussed in \cite{AToia:2014}, the \qpa\ calculated with
multiplicity based estimator (shown in Fig.\ref{fig:QpAhybrid} for
CL1, where centrality is based on the tracklets measured in
$|\eta|<1.4$) widely spread between centrality classes. They also exhibit a
negative slope in \pt, mostly in periphearl events, due to the jet
veto bias, as jet contribution increases with \pt.  The \qpa\ compared
to G-PYTHIA, a toy MC which couples Pythia to a p-Pb Glauber MC, show
a good agreement, everywhere in 80-100\%, and in general at high-\pt,
demonstrating that the proper scaling for high-\pt\ particle
production is an incoherent superposition of \pp\ collisions, provided
that the biased introduced by the centrality selection is properly
taken into account,as eg in G-PYTHIA. For ZNA centrality classes,
while no bias is expected on the multiplicity or high-Q$^2$ processes
is expected and indeed the classes present spectra which are much more
similar to each other, the absolute values of the spectra at high-\pt
indicate the presence of a bias, not due to the event selection but
because of inaccurate \Ncoll\ values calculated with the SNM.

With the hybrid method, using either the assumption on mid-rapidity
multiplicity proportional to \Npart, or forward multiplicity
proportional to \Nparttar, shown in Fig.~\ref{fig:QpAhybrid}, result in
consistent \qpa, also consistent with one for all centrality classes,
also observed for MB collisions,
indicating the absence of initial state effects. The observed Cronin
enhancement is stronger in central collisions and nearly absent in
peripheral collisions. The enhancement is also weaker at LHC energies
compared to RHIC energies. 
%%%%%%%%%%%%%%%%%%%%%
%%%%%%%%%%%%%%%%%%%%%

\subsection{Charged particle density}
\begin{figure}[t!f]
 \centering 
 \includegraphics[width=0.55\textwidth]{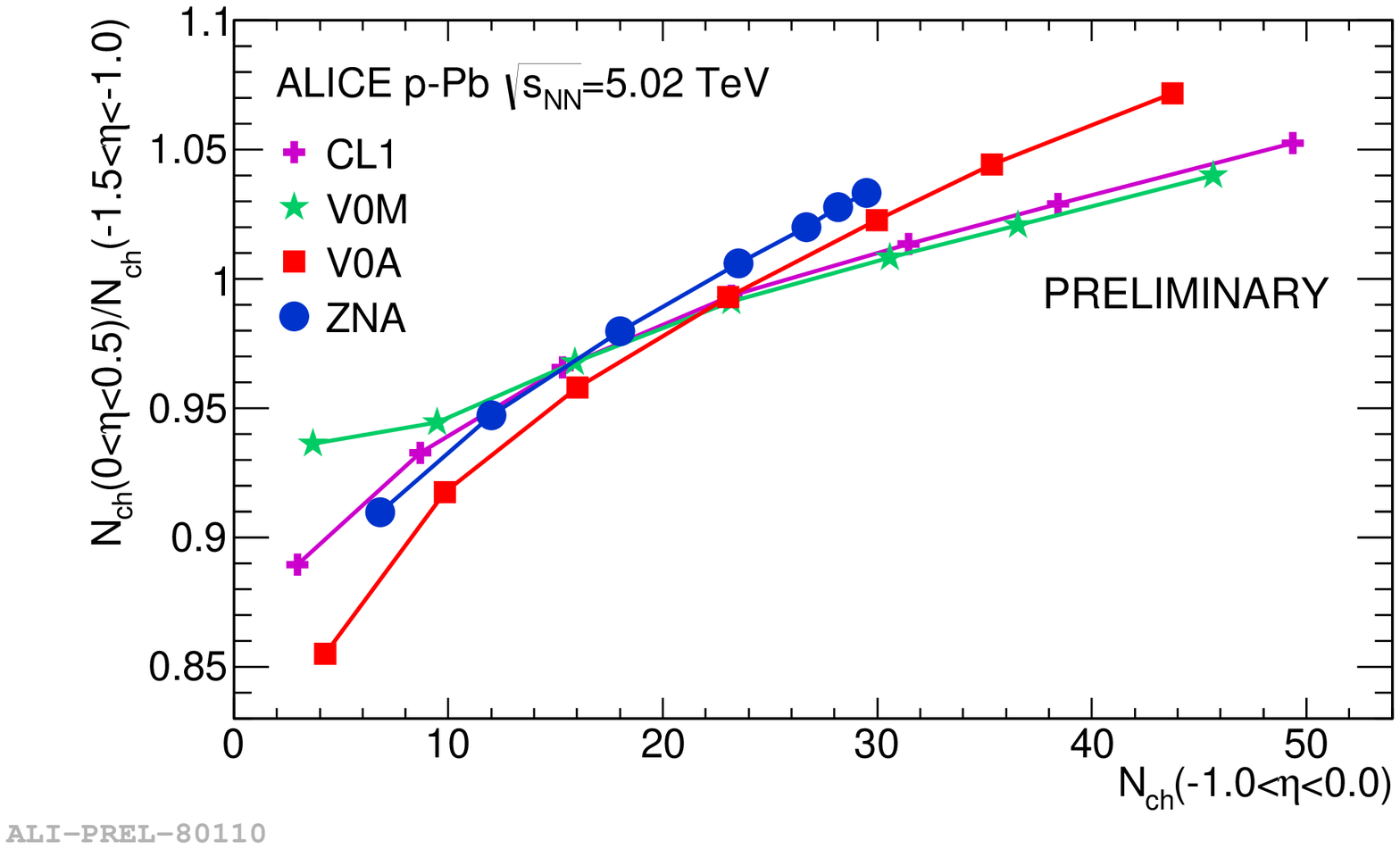}
 \includegraphics[width=0.4\textwidth]{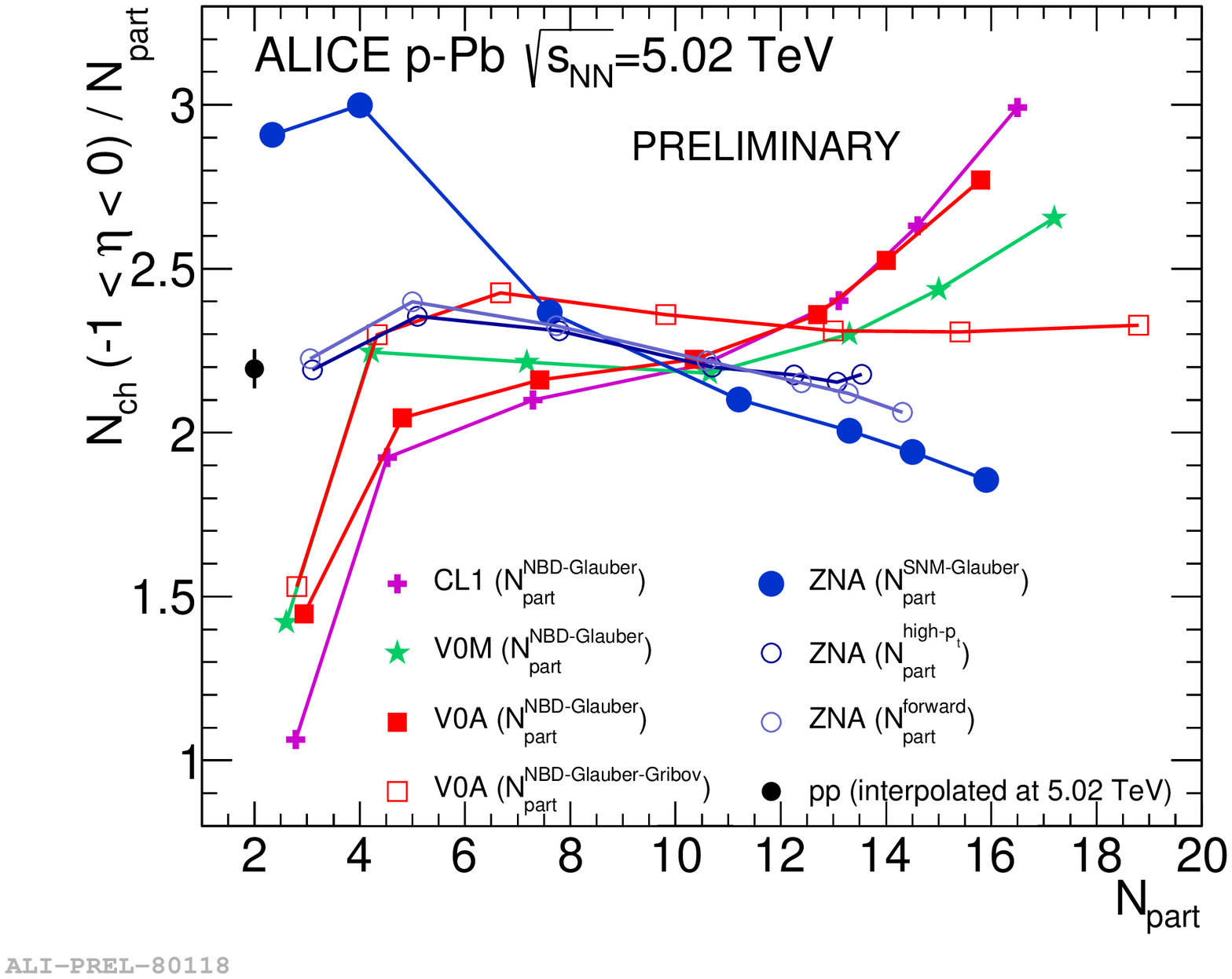}
 \caption{Left: Asymmetry of particle yield, as a function of the
   pseudorapidity density at mid-rapidity for various centrality
   classes and estimators. Right: Pseudorapidity density of charged
   particles at mid-rapidity per participant as a function of
   \Npart\ for various centrality estimators.
\label{fig:dndetaNpart2}}
\end{figure}

Charged particle density is also studied as a function of $\eta$, for
different centrality classes, with different estimators. In peripheral
collisions the shape of the distribution is almost fully symmetric and
resembles what is seen in proton-proton collisions, while in central
it becomes progressively more asymmetric, with an increasing excess of
particles produced in the direction of the Pb beam. We have quantified
the evolution plotting the asymmetry between the proton and lead peak
regions, as a function of the yield around the center of mass (see
Fig.~\ref{fig:dndetaNpart2}, left): the increase of the asymmetry is
different for the different estimators.  Fig.~\ref{fig:dndetaNpart2}
right shows \Nch\ at mid-rapidity divided by \Npart\ as a function of
\Npart\ for various centrality estimators. For Multiplicity-based
estimators (CL1, V0M, V0A) the charged particle density at mid
rapidity increases more than linearly, as a consequence of the strong
multiplicity bias. This trend is absent when \Npart\ are calculated
with the Glauber-Gribov model, which shows a relatively constant
behavior, with the exception of the most peripheral point.  For ZNA,
there is a clear sign of saturation above \Npart\ = 10, due to the
saturation of forward neutron emission. None of these curves points
towards the \pp\ data point. In contrast, the results obtained with
the hybrid method, using either \Nparttar-scaling at forward rapidity
or \Ncoll-scaling for high-\pt\ particles give very similar trends,
and show a nearly perfect scaling with \Npart, which naturally reaches
the \pp\ point. This indicates the sensitivity of the \Npart-scaling
behavior to the Glauber modeling, and the importance of the
fluctuations of the nucleon-nucleon collisions themselves.

\section{Conclusions}
Multiplicity Estimators induce a bias on the hardness of the pN
collisions.  When using them to calculate centrality-dependent \qpa,
one must include the full dynamical bias.  However, using the
centrality from the ZNA estimator and our assumptions on particle
scaling, an approximate independence of the multiplicity measured at
mid-rapitity on the number of participating nucleons is observed.
Furthermore, at high-\pt\ the \pPb\ spectra are found to be consistent
with the pp spectra scaled by the number of binary nucleon--nucleon
collisions for all centrality classes. Our findings put strong
constraints on the description of particle production in high-energy
nuclear collisions.

\end{document}